\documentclass[prl,twocolumn,tightenlines,superscriptaddress,nofootinbib,showpacs]{revtex4}
\usepackage{amssymb,latexsym}
\usepackage{amsmath,amsbsy,bbm}
\usepackage{epsfig,bm}
\usepackage{graphicx,comment}
\usepackage{color}
\newcommand{\p}{\partial}

\begin{document}

\title{Very High Precision Determination of Low-Energy Parameters: \\
The 2-d Heisenberg Quantum Antiferromagnet as a Test Case}

\author{F.-J. Jiang}
\email[]{fjjiang@ntnu.edu.tw}
\affiliation{Department of Physics, National Taiwan Normal University, 88, 
Sec.\ 4, Ting-Chou Rd., Taipei 116,Taiwan}
\author{U.-J. Wiese}
\affiliation{Albert Einstein Center for Fundamental Physics, Institute for 
Theoretical Physics, Sidlerstrasse 5, CH-3012 Bern, Switzerland}

\vspace{-2cm}

\begin{abstract}

The 2-d spin $\frac{1}{2}$ Heisenberg antiferromagnet with exchange coupling $J$
is investigated on a periodic square lattice of spacing $a$ at very small 
temperatures using the loop-cluster algorithm. 
Monte Carlo data for the staggered and uniform susceptibilities are compared 
with analytic results obtained in the systematic low-energy effective field 
theory for the staggered magnetization order parameter. The low-energy 
parameters of the effective theory, i.e.\ the staggered magnetization density 
${\cal M}_s = 0.30743(1)/a^2$, the spin stiffness $\rho_s = 0.18081(11) J$, and
the spin wave velocity $c = 1.6586(3) J a$ are determined with very high 
precision. Our study may serve as a test case for the comparison of lattice QCD
Monte Carlo data with analytic predictions of the chiral effective theory for 
pions and nucleons, which is vital for the quantitative understanding of the 
strong interaction at low energies.
\end{abstract}

\pacs{12.39.Fe, 75.10.Jm, 02.70.Ss, 11.30.Qc}

\maketitle

{\bf Introduction} ---
Partly motivated by the relation of antiferromagnetism to high-temperature 
superconductivity, during the past twenty years quantum spin models, such as the
spin $\frac{1}{2}$ Heisenberg antiferromagnet on the square lattice, have been 
studied in great detail. Since this system is strongly coupled, numerical 
simulations play an important role in its quantitative analysis. In this way, 
it has been shown that the $SU(2)_s$ spin symmetry breaks down spontaneously to 
its $U(1)_s$ subgroup at zero temperature. As a result, massless Goldstone 
bosons --- the antiferromagnetic magnons --- dominate the low-energy physics.
The magnon dynamics can be described quantitatively using a low-energy 
effective field theory for the staggered magnetization order parameter 
\cite{Cha89,Neu89,Has90,Has91,Has93}.  Low-energy phenomena can then be 
investigated analytically, order by order in a systematic derivative expansion.

Systematic effective field theories also play an important role in the 
low-energy physics of the strong interaction. On the one hand, lattice QCD 
describes the underlying dynamics of quarks and gluons beyond perturbation
theory, but can only be investigated by very large scale
Monte Carlo calculations. On the other hand, chiral perturbation theory, the
systematic low-energy effective field theory for pions --- the pseudo-Goldstone
bosons of the spontaneously broken $SU(2)_L \times SU(2)_R$ chiral symmetry of
QCD --- has been investigated analytically in great detail. The predictions of
the effective theory depend on a number of low-energy parameters, including the
pion decay constant, the chiral condensate, as well as the higher-order
Gasser-Leutwyler coefficients. For the quantitative understanding of the strong
interaction at low energies, it is of central importance to accurately determine
the values of the low-energy parameters by comparison of lattice QCD Monte Carlo
data with analytic chiral perturbation theory predictions. In recent years, 
there has been substantial progress in this direction, and the leading-order 
low-energy parameters have been determined with a few percent accuracy. 
Extending this to the higher-order low-energy parameters, as well as
reaching higher precision while keeping complete control of systematic errors
will be a major challenge for lattice QCD in the near future.

The 2-d spin $\frac{1}{2}$ Heisenberg antiferromagnet can serve as an ideal
test case, in which the interplay between high-precision numerical simulations
of the underlying microscopic system and high-order calculations in the 
corresponding systematic low-energy effective field theory can be investigated
quantitatively. In contrast to lattice QCD which is much more complicated, the
Heisenberg model can be simulated with very efficient methods, and has been 
investigated in several high-accuracy numerical studies
\cite{Wie94,Bea96,San97,Bea98,San99,Kim00,Wan05,San08,Wen08,Jia09,Ger09}
The first very precise determination of the low-energy constants of the 2-d 
spin $\frac{1}{2}$ Heisenberg antiferromagnet was performed in 
\cite{Wie94} using the loop-cluster algorithm \cite{Eve93}. This study was 
based on a cubical space-time geometry, for which the inverse temperature 
$\beta = 1/T$, which determines the extent of Euclidean time, is compatible 
with the spatial size $L$, i.e.\ $\beta c \approx L$. By comparison of the 
Monte Carlo data with analytic 2-loop results of Hasenfratz and Niedermayer, 
obtained in the systematic low-energy effective field theory for the staggered 
magnetization order parameter \cite{Has93}, the staggered magnetization density 
${\cal M}_s$, the spin stiffness $\rho_s$, and the spin wave velocity $c$ were 
determined as ${\cal M}_s = 0.3074(4)/a^2$, $\rho_s = 0.186(4) J$, and 
$c = 1.68(1) J a$. A few years later, the development of the continuous-time 
simulation technique \cite{Bea96} enabled numerical investigations of quantum 
spin models at very low temperatures. This allowed a comparison of Monte Carlo 
data with analytic 1-loop results in the cylindrical space-time regime at very 
low temperatures $\beta c \gg L$, which led to ${\cal M}_s = 0.3083(2)/a^2$, 
$\rho_s = 0.185(2) J$, and $c = 1.68(1) J a$, in statistical agreement with the 
results obtained in the cubical space-time regime. The fit in the cylindrical 
regime required an incorporation of 2-loop corrections with adjustable 
pre\-fac\-tors, because these effects had not been determined analytically at 
that time. Recently, Niedermayer and Weiermann have closed this gap by performing the
corresponding analytic 2-loop calculation in the effective theory \cite{Nie10}.
Slab-like space-time geometries with $L \gg \beta c$ have been investigated in
\cite{Bea98}. In that study, using finite-size scaling, very long spatial 
correlation lengths up to 350000 lattice spacings have been investigated. A
combined fit of Monte Carlo data in the cubical, cylindrical, and slab 
geometries then gave ${\cal M}_s = 0.30797(3)/a^2$, $\rho_s = 0.1800(5) J$, and 
$c = 1.657(2) J a$. In a recent study using a zero-temperature valence-bond
projector method, Sandvik and Evertz obtained the very accurate result
${\cal M}_s = 0.30743(1)/a^2$. Although the discrepancy between these two 
results for ${\cal M}_s$ is at the per mille level, it is statistically
significant. Indeed, in a high-precision analysis of the constraint effective 
potential of the staggered magnetization, which relied on 2-loop predictions
in the effective theory by G\"ockeler and Leutwyler \cite{Goe91,Goe91a}, we
suspected that the previously obtained estimates ${\cal M}_s = 0.3083(2)/a^2$
\cite{Bea96} and ${\cal M}_s = 0.30797(3)/a^2$ \cite{Bea98}, which were 
dominated by Monte Carlo data in the cylindrical regime, are afflicted by an 
underestimated systematic error resulting from a truncation of the Seeley
expansion described in \cite{Has93}. In this paper, we return to the cylindrical
regime and clarify the discrepancy. This will result in a confirmation of the 
value ${\cal M}_s = 0.30743(1)/a^2$ obtained in \cite{San08}, as well as in a
determination of $\rho_s = 0.18081(11) J$ and $c = 1.6586(3) J a$ with
unprecedented precision. In the cubical regime we determine $c$ by tuning
$\beta$ until the squares of the spatial and temporal winding numbers become
identical. In this way, the uncertainties of both ${\cal M}_s$ and $\rho_s$ 
resulting from the fits are drastically reduced.

While reaching fractions of a per mille precision for the low-energy parameters 
may seem unnecessary from a condensed matter physics perspective, it is 
reassuring for the ongoing efforts to combine lattice QCD with chiral 
perturbation theory in order to accurately determine the fundamental low-energy 
parameters of the strong interaction. Using the 2-d Heisenberg model as a test 
case, our analysis demonstrates that very precise Monte Carlo data combined with
2-loop effective field theory predictions for a variety of physical quantities 
indeed leads to a completely consistent very high precision determination of the
fundamental low-energy parameters. \\

{\bf Microscopic Model and Corresponding Observables} ---
The spin $\frac{1}{2}$ Heisenberg model considered in this study is defined by 
the Hamilton operator
\begin{eqnarray}
\label{hamilton}
H = \sum_x J \Big[\,\vec S_x \cdot \vec S_{x+\hat{1}}+ 
\vec S_x \cdot \vec S_{x+\hat{2}}\,\Big],
\end{eqnarray}
where $\hat{1}$ and $\hat{2}$ refer to the two spatial unit-vectors. Further, 
$J$ in eq.(\ref{hamilton}) is the antiferromagnetic exchange coupling. A 
physical quantity of central interest is the staggered susceptibility
\begin{equation}
\label{defstagg}
\chi_s = \frac{1}{L^2} \int_0^\beta dt \ \frac{1}{Z} 
\mbox{Tr}[M^3_s(0) M^3_s(t) \exp(- \beta H)].
\end{equation}
Here $Z = \mbox{Tr}\exp(- \beta H)$ is the canonical partition function. The 
staggered magnetization order parameter is defined as 
$\vec M_s = \sum_x (-1)^{x_1+x_2} \vec S_x$. Another relevant quantity is the 
uniform susceptibility
\begin{equation}
\label{defuniform}
\chi_u = \frac{1}{L^2} \int_0^\beta dt \ \frac{1}{Z} \mbox{Tr}[M^3(0) M^3(t)
\exp(- \beta H)].
\end{equation}
Here $\vec{M} = \sum_x \vec S_x$ is the uniform magnetization. Both $\chi_s$ and
$\chi_u$ can be measured very accurately with the loop-cluster algorithm using 
improved estimators \cite{Wie94}. In particular, in the multi-cluster version of
the algorithm the staggered susceptibility is given in terms of the cluster 
sizes $|{\cal C}|$ as $\chi_s = \frac{1}{\beta L^2} 
\left\langle \sum_{\cal C} |{\cal C}|^2 \right\rangle$. Similarly, the uniform 
susceptibility $\chi_u = \frac{\beta}{ L^2 } \left\langle W_t^2 \right\rangle =
\frac{\beta}{L^2} \left\langle \sum_{\cal C} W_t({\cal C})^2 \right\rangle$
is given in terms of the temporal winding number 
$W_t = \sum_{\cal C} W_t({\cal C})$ which is the sum of winding numbers
$W_t({\cal C})$ of the loop-clusters ${\cal C}$ around the Euclidean time 
direction. Similarly, the spatial winding numbers are defined by 
$W_i = \sum_{\cal C} W_i({\cal C})$ with $i \in \{1,2\}$. \\

{\bf Low-Energy Effective Theory for Magnons} ---
Due to the spontaneous breaking of the $SU(2)_s$ spin symmetry down to its 
$U(1)_s$ subgroup, the low-energy physics of an antiferromagnet is governed by
two massless Goldstone bosons, the magnons. A systematic low-energy effective 
field theory for magnons was developed in \cite{Cha89,Neu89,Has90,Has91}. The 
staggered magnetization of an antiferromagnet is described by a unit-vector 
field $\vec{e}(x)$ that takes values in the coset space $SU(2)_s/U(1)_s = S^2$, 
i.e.\ $\vec e(x) = \big(e_1(x),e_2(x),e_3(x)\big)$ with $\vec e(x)^2 = 1$.
Here $x = (x_1,x_2,t)$ denotes a point in (2+1)-dimensional space-time. To 
leading order, the Euclidean magnon low-energy effective action takes the form
\begin{eqnarray}
\label{action}
S[\vec e\,]&=&\int^L_0 dx_1 \int^L_0 dx_2 \int^\beta_0 \ dt \nonumber \\  
&\times&\frac{\rho_s}{2} \left(\p_1 \vec e \cdot \p_1 \vec e + 
\p_2 \vec e \cdot \p_2 \vec e + \frac{1}{c^2} \p_t \vec e \cdot \p_t \vec e
\right),
\end{eqnarray}
where $t$ refers to the Euclidean time-direction. It should be noted that the 
effective field theory described by eq.(\ref{action}) is valid as long as the 
conditions $L \rho_s \gg 1$ and $\beta c \rho_s \gg 1$ are satisfied. As 
demonstrated in \cite{Wie94}, once these conditions are satisfied, the 
low-energy physics of the underlying microscopic model can be captured 
quantitatively by the effective field theory. Using the systematic effective 
theory, detailed calculations of a variety of physical quantities including
2-loop corrections have been carried out in \cite{Has93}. Here we only quote 
the results that are relevant to our study. The aspect ratio of a spatially 
quadratic space-time box of spatial size $L$ is characterized by 
$l = (\beta c/L)^{1/3}\,,$ which distinguishes cubical space-time
volumes with $\beta c \approx L$ (known as the $\epsilon$-regime in QCD) from 
cylindrical ones with $\beta c \gg L$ (the so-called $\delta$-regime in QCD). 
In the cubical regime, the volume- and temperature-dependence of the staggered 
susceptibility is given by
\begin{eqnarray}
\label{chiscube}
\chi_s&=&\frac{{\cal M}_s^2 L^2 \beta}{3} 
\left\{1 + 2 \frac{c}{\rho_s L l} \beta_1(l) \right. \nonumber \\
&&+\left.\left(\frac{c}{\rho_s L l}\right)^2 \left[\beta_1(l)^2 + 
3 \beta_2(l)\right] + O\left(\frac{1}{L^3}\right) \right\},
\end{eqnarray}
while the uniform susceptibility takes the form
\begin{eqnarray}
\label{chiucube}
\chi_u&=&\frac{2 \rho_s}{3 c^2} 
\left\{1 + \frac{1}{3} \frac{c}{\rho_s L l} \widetilde\beta_1(l) +
\frac{1}{3} \left(\frac{c}{\rho_s L l}\right)^2 \right. \nonumber \\
&\times&\left.\left[\widetilde\beta_2(l) - \frac{1}{3} \widetilde\beta_1(l)^2 - 
6 \psi(l)\right] + O\left(\frac{1}{L^3}\right) \right\}.
\end{eqnarray}
In eqs.(\ref{chiscube}) and (\ref{chiucube}), the functions $\beta_i(l)$, 
$\widetilde\beta_i(l)$, and $\psi(l)$, which only depend on $l$, are known shape
coefficients of the space-time box defined in \cite{Has93}. Finally, in the 
cylindrical regime,  when the condition $L^2\rho_s/(\beta c^2) \ll 1$ is 
satisfied, the volume-dependence of the staggered susceptibility is given by
\begin{eqnarray}
\label{chiscylin}
\chi_s&=&\frac{2}{3}\frac{{\cal M}^2_s \rho_s  L^4}{c^2} 
\left\{1 + 3a \frac{c}{\rho_s L} 
+ 3a^2 \left(\frac{c}{\rho_s L}\right)^2 \right. \nonumber \\
&-&\left. b \left(\frac{c}{\rho_s L}\right)^2 + O\left(\frac{1}{L^3}\right) 
\right\}\,,
\end{eqnarray}
where $a = 0.3103732207$ and $b = 0.0004304999$ \cite{Nie10}. It should be noted
that $\chi_s$ given in eq.(\ref{chiscylin}) is temperature-independent. \\

\begin{figure}
\begin{center}
\vskip0.75cm
\includegraphics[width=0.48\textwidth]{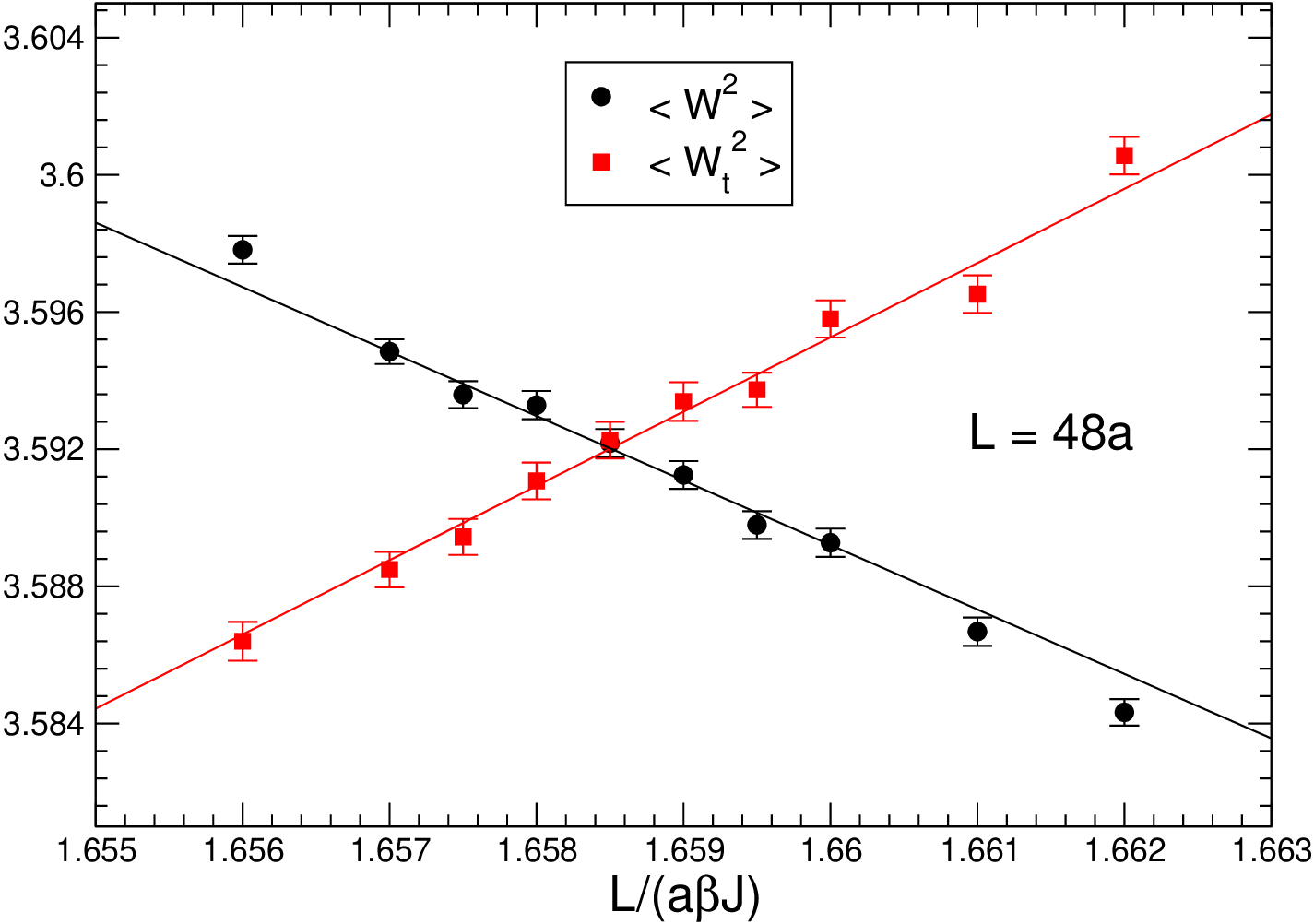}
\end{center}
\vskip-0.5cm
\caption{The determination of $c$ using the squares of spatial and temporal 
winding numbers at $L = 48a$.}
\label{fig1}
\end{figure}

{\bf Determination of Low-Energy Parameters} --- In order to determine the 
low-energy parameters ${\cal M}_s$, $\rho_s$, and $c$ for the spin $\frac{1}{2}$
Heisenberg model on the square lattice, we have performed large-scale 
simulations for various inverse temperatures $\beta$ and box sizes $L$. We 
determine $c$ using the idea proposed in \cite{Jia10}: for a fixed box size 
$L$, we vary $\beta$ until the condition $\langle W_t^2 \rangle = 
\frac{1}{2}(\langle W_1^2\rangle + \langle W_2^2\rangle)$ is satisfied. The
spin wave velocity then results as $c = L/\beta$. Using this method, we obtain
$c = 1.6586(3)Ja$ (see figure \ref{fig1}).
This value is obtained by performing a weighted average over the values of $c$ 
listed in table I, which are extracted in volumes ranging from $L = 24 a$ to 
$L = 64 a$.
\begin{table}
\begin{center}
\begin{tabular}{|c|c|}
\hline
$L/a$ & $c$ \\
\hline
\hline
24 & 1.6589(6) \\
\hline
32 & 1.6586(5) \\
\hline
48 & 1.6585(5) \\
\hline
64 & 1.6585(5) \\
\hline
\end{tabular}
\end{center}
\caption{Values of $c = L/\beta$ extracted for different lattice sizes 
$L/a$ by tuning $\beta$ such that the average squares of the spatial and 
temporal winding numbers are the same.}
\end{table}
It should be noted that the above value 
of $c$ is consistent with the one quoted in \cite{Bea98}, but the statistical
error is reduced by a factor of 7. In principle, using this method one could 
obtain an even more precise estimate of $c$. After obtaining this very accurate 
value of $c$, we carry out further large scale simulations in the cubical regime
with $\beta c \approx L$. Using $c = 1.6586(3)Ja$ and performing a combined fit 
of the Monte Carlo data for $\chi_s$ and $\chi_u$ in the cubical regime to 
eqs.(\ref{chiscube}) and (\ref{chiucube}), we arrive at 
${\cal M}_s = 0.30743(1)/a^2$ and $\rho_s = 0.18081(11)J$ with
$\chi^2/{\text{d.o.f.}} \approx 1$. 
\begin{figure}
\begin{center}
\vskip0.5cm
\vbox{
\includegraphics[width=0.48\textwidth]{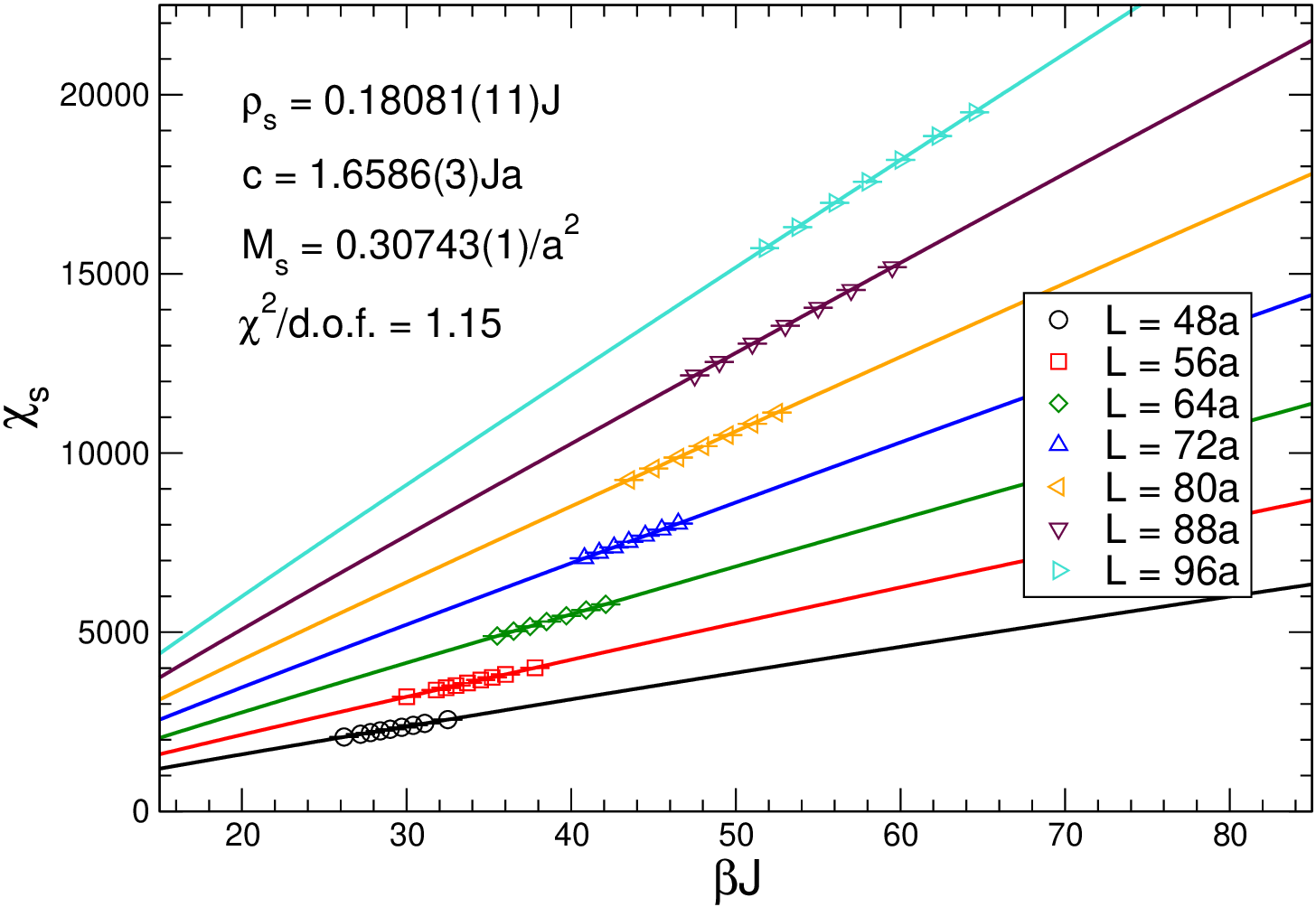}\vskip0.85cm
\includegraphics[width=0.48\textwidth]{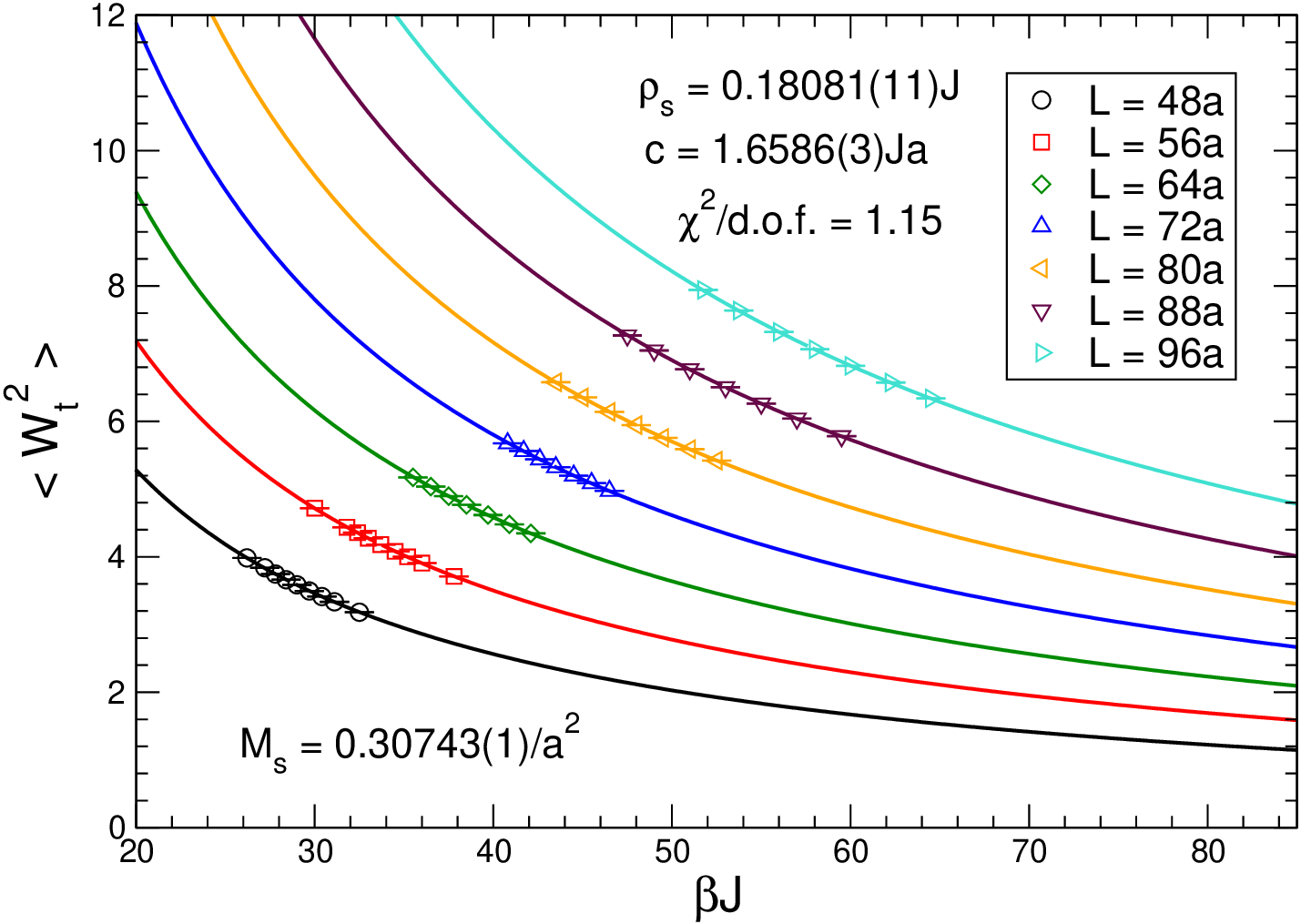}
}
\end{center}
\vskip-0.5cm
\caption{Fits of $\chi_s$ and $\langle W_t^2\rangle$ (and thus $\chi_u$) to 
their predicted behavior in magnon chiral perturbation theory. For better 
visibility, some data used in the fits are omitted in the figure.}
\label{fig2}
\end{figure} 
Figure \ref{fig2} illustrates the results of
the fit. The main contribution to the uncertainties of ${\cal M}_s$ and $\rho_s$
results from the error of $c$ that enters the fit. Hence, with a more precise 
estimate of $c$, one could even further improve the accuracy of ${\cal M}_s$ 
and $\rho_s$. The values we obtain for $\rho_s$ and $c$ are more accurate than 
earlier estimates of these low-energy parameters. It should be noted that the
value obtained for ${\cal M}_s$ is consistent with the one of \cite{San08}, and 
the statistical error is the same in both cases.

Next we simulate the model in the cylindrical regime where the condition 
$\beta c \gg L$ is satisfied. Since a main motivation of our study is to 
clarify the discrepancy between the values of ${\cal M}_s$ presented in 
\cite{Bea98} and \cite{San08}, and the accuracy we must reach is hence below 
the per mille level, we adopt the following strategy. First, we note that at 
very low temperatures $\chi_s$ becomes temperature-independent. In order to 
avoid underestimating the systematic errors in an extrapolation to zero 
temperature, we simulate at sufficiently low temperatures so that $\chi_s$  
becomes independent of $\beta$ within error bars. Second, it should be noted 
that Monte Carlo data for 
$\chi_s$ in both the cubical \cite{Wie94} and the cylindrical regime 
\cite{Bea96} were used for obtaining the value of ${\cal M}_s$ quoted in 
\cite{Bea98}. Since both $\rho_s$ and $c$ obtained in the cubical regime
are consistent with the corresponding results in the cylindrical regime, one 
may conclude that the overestimation of ${\cal M}_s = 0.30797(3)/a^2$ presented 
in \cite{Bea98} is due to the cylindrical regime data for $\chi_s$. In order to 
minimize statistical correlations between $\chi_s$ and $\chi_u$, in our 
fitting strategy we use only cylindrical regime data for $\chi_s$ and cubical 
regime data for $\chi_u$. Applying these strategies and using $c = 1.6586(3)Ja$,
we arrive at ${\cal M}_s = 0.30746(4)/a^2$ and $\rho_s = 0.18081(11)J$
(see figure 3). 
\begin{figure}
\begin{center}
\vskip0.5cm
\includegraphics[width=0.48\textwidth]{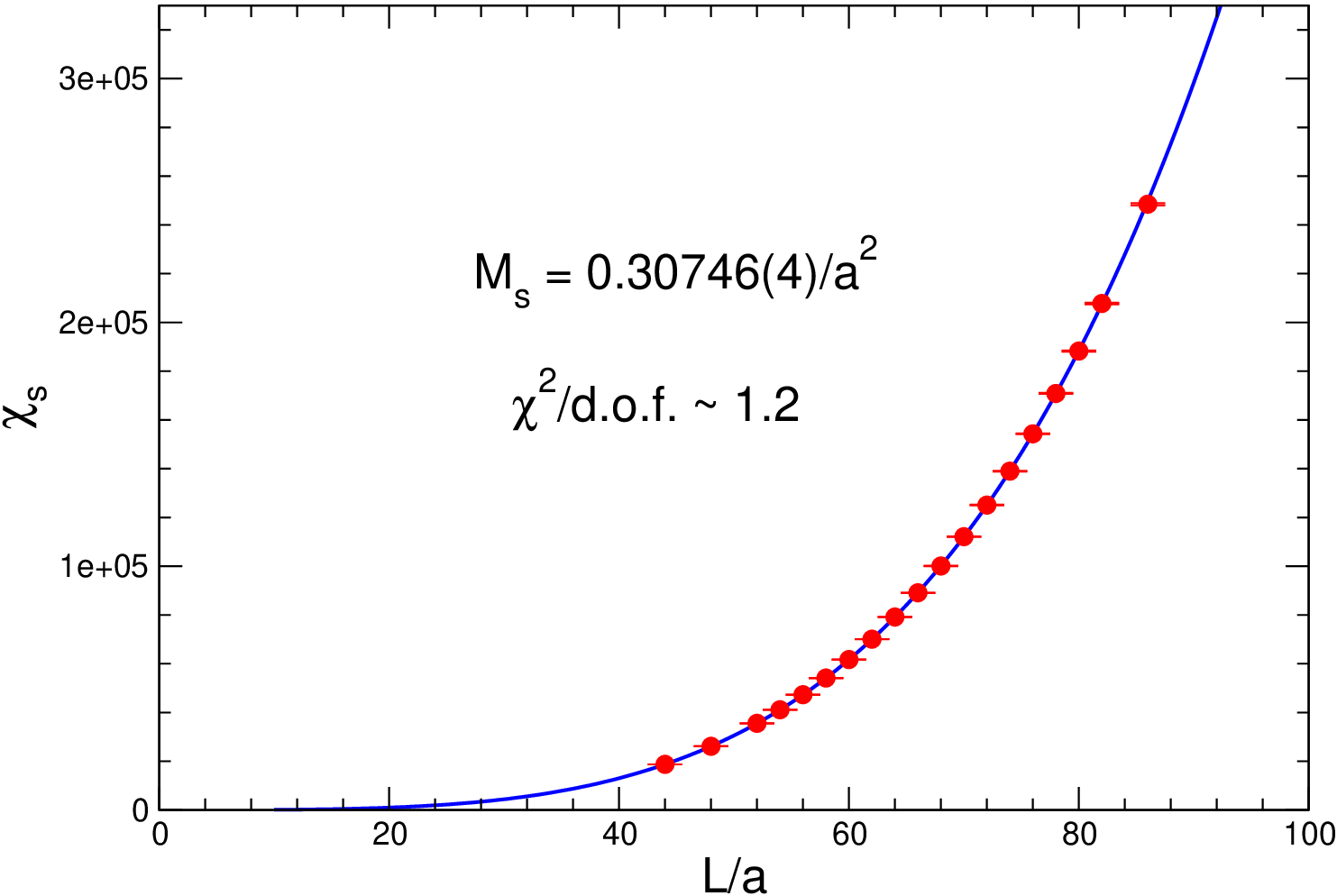}\vskip0.85cm
\end{center}
\vskip-1cm
\caption{Fit of Monte Carlo data for $\chi_s$ in the cylindrical regime to 
their chiral perturbation theory prediction.}
\label{fig3}
\end{figure}   
It should be noted that this value of ${\cal M}_s$, which we obtain in the 
cylindrical regime, is consistent with the value determined in the cubical 
regime. It is also consistent with the most accurate result for ${\cal M}_s$
that was previously obtained \cite{San08}.

In the cylindrical regime, simulating larger lattices is necessary in order to
reach the same accuracy for ${\cal M}_s$ as the one obtained in \cite{San08}.
This demonstrates the advantage of finite-temperature simulations: applying the 
effective field theory predictions to finite-temperature data, which can be 
obtained with a moderate computational effort, one achieves a very precise 
numerical value for ${\cal M}_s$. Using $c = 1.6586(3)Ja$, we arrive at 
${\cal M}_s = 0.30743(1)/a^2$ and $\rho_s = 0.18081(10)J$ from a combined fit 
including all available data points. The accuracy of these low-energy constants 
is not improved compared to those obtained in the cubical regime alone. This is 
reasonable since only a few more data points are included in the new fit.

Finally, we would like to clarify possible reasons for the overestimation of 
${\cal M}_s = 0.30797(3)/a^2$ obtained in \cite{Bea98}. Because of the 
consistency of both $\rho_s$ and $c$ obtained in the cubical and cylindrical 
regimes \cite{Wie94,Bea96}, one concludes that the slight overestimation of 
${\cal M}_s$ in \cite{Bea98} is due to the cylindrical regime data for $\chi_s$.
In particular, in order to employ eq.(\ref{chiscylin}) to determine 
${\cal M}_s$, in \cite{Bea96} a Seeley expansion has been performed in 
order to extrapolate the finite-temperature $\chi_s$ data to their corresponding
zero-temperature limit. However, terminating the Seeley series is a subtle 
matter. Hence, the 
Seeley extrapolation may lead to an underestimated systematic error  if the data
are outside the window in which such an extrapolation is justified. Because of 
this, instead of repeating the analysis performed in \cite{Bea96}, we adopt
another strategy. Specifically, we again simulate the model with box sizes 
$L/a = 10, 12, 14,\dots,20$ at sufficiently low temperatures so that the 
$\chi_s$ data we obtain are independent of $\beta$. A combined fit of these 
newly obtained data for $\chi_s$ at very low temperature to eq.(\ref{chiscylin})
and the $\chi_u$ data we obtained earlier in the cubical regime to 
eq.(\ref{chiucube}) yields ${\cal M}_s = 0.3070(2)/a^2$, $\rho_s = 0.182(2) J$ 
and $c = 1.66(1) Ja$ with $\chi^2/{\text{d.o.f.}} \approx 1.3$. While the 
obtained value ${\cal M}_s = 0.3070(2)/a^2$ is slightly below 
${\cal M}_s = 0.30743(1)/a^2$, the numerical values of the low-energy parameters
that we just obtained are indeed consistent with those found in the cubical 
regime calculations \cite{Wie94}. Therefore we conclude that the overestimation 
of ${\cal M}_s$ in \cite{Bea98} is most likely due to an underestimated 
systematic error of $\chi_s$ related to the termination of the Seeley 
expansion used in \cite{Bea96}. \\

{\bf Conclusions} --- In this letter, we have revisited the spin $\frac{1}{2}$ 
Heisenberg model on the square lattice. In particular, we have refined the 
numerical values of the corresponding low-energy parameters, namely the 
staggered magnetization density ${\cal M}_s$, the spin stiffness $\rho_s$, and 
the spin wave velocity $c$. The spin wave velocity is determined to very high 
accuracy using the squares of spatial and temporal winding numbers. Remarkably, 
using $c = 1.6586(3)Ja$ together with 2-loop magnon chiral perturbation theory 
predictions for $\chi_s$ and $\chi_u$ in the cubical regime, we obtained a good 
fit of more than 150 data points to two analytic expressions with only two 
unknown parameters. Specifically, from the fit we obtain 
${\cal M}_s = 0.30743(1)/a^2$ and $\rho_s = 0.18081(11)J$ with 
$\chi^2/{\text{d.o.f.}} \approx 1$. The precision of 
${\cal M}_s = 0.30743(1)/a^2$
is comparable to the one in \cite{San08} which was obtained by an 
unconstrained polynomial fit using up to third or fourth powers of $1/L$.
Furthermore, the data used in \cite{San08} were obtained at much larger $L$ 
than those used in our study. Thanks to the accurate predictions of magnon 
chiral perturbation theory, it requires only moderate computing resources to 
reach fraction of a per mille accuracy for the low-energy parameters. This
is encouraging for QCD where such accuracy is mandatory to reach a sufficiently
precise determination of the fundamental low-energy parameters of the strong
interaction. If calculations in the chiral limit would become feasible, the 
experience gained in the Heisenberg model would suggest that the hyper-cubical
$\varepsilon$-regime would be best suited for extracting the low-energy 
parameters. We have also resolved the puzzle of the per mille level discrepancy 
between the two values for ${\cal M}_s$ presented in \cite{Bea98} and 
\cite{San08} by re-simulating the model in the cylindrical regime. Based on a 
combined fit of $\chi_s$ in the cylindrical regime and $\chi_u$ in the cubical 
regime, we arrived at ${\cal M}_s = 0.30746(4)/a^2$, which is consistent with
both the results of \cite{Wie94} and \cite{San08}. The consistency of the values
for ${\cal M}_s$ obtained in the cubical and in the cylindrical regime is 
particularly remarkable in view of the increased analytic 2-loop accuracy
in the cylindrical regime \cite{Nie10}, which demonstrates the quantitative 
correctness of magnon chiral perturbation theory in describing the low-energy 
physics of the underlying microscopic model. 
Finally, we concluded that the small discrepancy in the values for 
${\cal M}_s$ between \cite{Bea98} and \cite{San08} may be attributed to the 
termination of the Seeley expansion used in obtaining the former result.

We like to thank F.\ Niedermayer for very helpful discussions. Partial support 
from NSC and NCTS (North) as well as from the Schweizerischer Nationalfonds is 
acknowledged. The ``Albert Einstein Center for Fundamental Physics'' at Bern 
University is supported by the ``Innovations- und Kooperationsprojekt C-13'' of
the Schweizerische Universit\"atskonferenz (SUK/CRUS).

\end{document}